\documentclass[journal,draftcls,onecolumn,12pt,twoside]{IEEEtranTCOM}
\normalsize

\usepackage{graphicx}

\ifCLASSINFOpdf
\else
\fi

\ifCLASSINFOpdf
\else
\usepackage[dvips]{graphicx}
\fi
\usepackage{url}

\usepackage{xcolor}

\usepackage{amssymb}  
\usepackage{amsmath}  
\usepackage{latexsym} 
\usepackage{subcaption}
\newtheorem{thm}{Theorem}[section]

\newtheorem{cor}[thm]{Corollary}
\newtheorem{definition}[thm]{Definition}
\newtheorem{rem}[thm]{Remark}
\newtheorem{notation}[thm]{Notation}

\newcommand{\red}[1]{\textcolor{black}{#1}}
\newcommand{\magenta}[1]{\textcolor{black}{#1}}

\DeclareMathOperator{\Res}{Res}

\begin{document}
%
\title{\red{Nonlinear Fourier spectrum characterization of time-limited signals}
}
%
%
%

\author{Dmitry Shepelsky, Anastasiia Vasylchenkova, Jaroslaw E. Prilepsky, and Iryna Karpenko
	\thanks{This work was supported in part by the Leverhulm Trust Grant RP-2018-063 and Erasmus+ mobility exchange programme.}
	\thanks{D. Shepelsky and I. Karpenko are with B. Verkin Institute for Low Temperature Physics and Engineering, Kharkiv, Ukraine (e-mails: shepelsky@yahoo.com  and 
		inic.karpenko@gmail.com).}
	\thanks{A. Vasylchenkova and J. E. Prilepsky are with the Aston Instotute of Photonic Technologies, Aston University, Birmingham, B4 7ET, UK  (e-mails: vasylcha \& y.prylepskiy1@aston.ac.uk).}}

%
%

\markboth{IEEE Transactions on Communications}%
{Submitted paper}
%



\maketitle

\begin{abstract}
Addressing the optical communication systems employing the nonlinear Fourier transform (NFT) for the data modulation/demodulation, we provide an explicit proof for the properties of the signals emerging in the so-called $b$-modulation method, the nonlinear signal modulation technique that provides explicit control over the signal extent. \red{We present details of the procedure and related rigorous mathematical proofs addressing the case where the time-domain profile corresponding to the $b$-modulated data has a limited duration, and when the bound states corresponding to specifically chosen discrete solitonic eigenvalues and norming constants, are also present. We also prove that the number of solitary modes that we can embed without violating the exact localisation of the time-domain profile, is actually infinite. Our theoretical findings are illustrated with numerical examples, where simple example waveforms are used for the $b$-coefficient, demonstrating the validity of the developed approach. We also demonstrate the influence of the bound states on the noise tolerance of the b-modulated system.}
\end{abstract}


%
\IEEEpeerreviewmaketitle

\section{Introduction}

\IEEEPARstart{I}{n} a multitude of different physical areas, and, notably, in fibre optics, the evolution of signals can often be well approximated by the nonlinear Schr\"{o}dinger equation (NLS) \cite{a10,ft07}. In particular, the latter serves as a leading order model that describes the propagation of light envelopes in fibre-optic communication channels under some simplifying conditions \cite{a10,tpl17}. The normalised lossless and noiseless NLS for the slow-varying complex electromagnetic field envelope function \red{$q(t,z)$}, where $z$ is the distance along the fibre and $t$ is the retarded time (in the fibre optics context), is given as follows
\begin{equation}\label{nlse}
iq_z + q_{tt} + 2|q|^2 q = 0 \, ,
\end{equation}
$i$ is the imaginary unit; for the explicit normalisations pertaining to single-mode optical fibres see e.g. \cite{tpl17}. The important property of the NLS (\ref{nlse}) is that it belongs to the class of so-called integrable equations, meaning that the initial-value problem for this equation can be solved by means of the inverse scattering technique \cite{ft07,zs72}, given some constraints on the ``initial conditions'', \red{$q(t,0)$} in our notations. The signal processing operations participating in this method are often referred to as NFT, and the multiplexing technique dealing with the nonlinear Fourier (NF)  domain data was coined nonlinear frequency division multiplexing \cite{yk14}. In a nutshell, the NFT maps the solution of NLS (i.e. our signal) onto the space of the complex-valued spectral parameter $k$, playing the role of a ``nonlinear frequency'', such that the NFT operation, Eq. (\ref{Z-Sh}) below, decomposes our space-time profile into the nonlinear modes evolving inside the NF domain. The nonlinear spectrum (i.e. the ``NFT image'') that corresponds to the initial profile with a finite first norm, $\red{q(t,0)} \in L^1(\mathbb R)$, contains, in the general case: 
\begin{enumerate}
	\item[(i)] two \red{complex} \textit{scattering coefficients} $a(k)$, $b(k)$ for $k \in \mathbb R$, describing the dispersive radiation components of our pulse;
	\item[(ii)]  the discrete (solitonic)  spectrum, consisting of two complex parameters for each discrete (soliton) mode: the eigenvalue $k_{j}$ and the respective spectral amplitude $c_j$.
\end{enumerate} 

\noindent Either the discrete or continuous part of the NF spectrum can be absent in some specific situations. See more explicit details in \cite{ft07,tpl17,zs72,yk14}. 

Insofar as the NF modes evolve linearly inside the NF domain, the NFT-based optical signal processing and the usage of the parameters of nonlinear modes as data carriers have been considered as an efficacious method for the nonlinearity mitigation in optical fibre links \cite{tpl17,yk14,lpt14,lab17}. The recently introduced $b$-modulation NFT technique \cite{w17}, operating with band-limited $b(k)$ profiles, has been aimed at resolving one of the principal challenges in the NFT-based communication: to attain explicit control over the temporal duration of the NFT-generated signals at the transmitter side \cite{lsbb18,lb18,yal18,gzl18}. The latter property allows us to pack our data better inside a given time-bandwidth volume and, thus, to reach \red{potentially} higher spectral efficiency numbers. In particular, the highest data rate reported so far for the NFT-based transmission method was achieved with a modified variant of $b$-modulation (in the dual-polarisation case) \cite{yal19} and has very recently been confirmed experimentally \cite{yla19}. In the case of $b$-modulation, we map our data on the function $b(k)$, which is chosen to be band-limited, and further adjust the function $a(k)$ accordingly, see the explanations and definitions below in Sec. \ref{sec:derivation}. Then, the ensuing signal $q(t)$, obtained through the inverse NFT operation, has a finite duration in the time domain \cite{w17}. It is exactly the latter feature that allows to get \red{a more efficient time-domain occupation} compared to ``conventional'' NFT-based systems employing the continuous NF spectrum modulation \cite{tpl17,lpt14}, \red{ the property which translates into a better system performance}, see the explicit comparison in \cite{gzl18}. \red{We, however, note that the $b$-modulation concept is not different in its set-up from other NFT-based methods. Thus, within this approach, each ``supersymbol'' generated from the $b$-modulated profile must be appended with zero-padding guard intervals in the time domain, which are equal in duration to the dispersion-induced memory. However, the duration of the ``supersymbol'' itself is shorter in comparison to the conventional continuous NF spectrum modulation where, typically, the time-domain waveforms develop ``a tail'', see \cite[Fig. 2 (d), (f)]{pdb14}.}

For the completeness of our exposition here, we note that in the original work by Wahls \cite{w17}, where the $b$-modulation concept was introduced, the problem of a complete characterization of $b(k)$ in the case when time-limited signals $q(t)$
support bound states (i.e. containing a non-zero discrete NF spectrum part), was formulated as an open question. In the follow-up study \cite{gzl18}, a necessary condition for the possibility to have bound states keeping the same $b(k)$ was stated and the analogy  with  the linear operator of the Lax pair representation for the Korteweg--de Vries (KdV) equation was mentioned. In the latter case, we deal with the bound states of the one-dimensional Schr\"odinger equation (written here for some function $\psi(t,k)$),
\[
\psi_{tt}+k^2 \psi = V(t)\psi.
\]

See works \cite{ap03, ac04}, where the necessary conditions for $b(k)$
to generate a  finitely supported $V(t)$, which serves in this context the role of $q(t)$ from the Zakharov-Shabat system (\ref{Z-Sh}), are discussed in the presence of non-zero discrete spectrum (bound states). \red{Portinari \cite{p78} pointed out that $b(\xi)$ being bandlimited is both necessary and sufficient for a finite support in the KdV case
if there are no bound states. 
For the NLS case (i.e., for the Zakharov-Shabat system as a spectral problem), 
the uniqueness of the determination of a time-limited $q(t)$ in (\ref{Z-Sh}) from $b(k)$ is discussed
in Ref.~\cite{s04} in the absence of bound states.
The arguments of Portinari were recently
carried over the NLS case in \cite{v18}, 
where the characterisation problem has also been addressed in the presence of eigenvalues,
by exploiting the spectral properties of  one-sided signals.}  
At this place we would like to emphasize that the recent \red{works \cite{vpc19,vpc19-1} propose to put an arbitrary additional solitary mode (or an ensemble of solitary components)} atop the $b$-modulated profile, but to keep the width of the solitary component in the time domain sufficiently thin. Such a composition ensures that the considerable portion of the \red{overall resulting} signal does not spread beyond the initial extent of the $b$-modulation-generated profile. Importantly, this approach \textit{does not provide truly localised signals}, and in our current study, we require a strict localisation of the time-domain signal, similarly to the initial definition of $b$-modulation \cite{w17}. \red{Finally, we note that the results of Brenne and Skaar in \cite{bs03} can be viewed as the discrete-time analogue of the results given in our paper.}

In this paper, we analyse $b$-modulated signals
making an emphasis on the sufficiency aspect. Namely, applying the Riemann--Hilbert
approach \cite{yk14,ks18,kvs18} for solving the inverse scattering problem for the Zakharov-Shabat system
(\ref{Z-Sh}), we \emph{characterize} time-limited signals having the same scattering 
coefficient $b(k)$  showing, in particular,  that it is possible to include the discrete nonlinear spectral components (solitons) into the $b$-modulation without violating the complete localisation of the respective time-domain profile. \red{In the end, it is noteworthy that whereas the inclusion of additional solitary components can increase the overall power of the signal and, thus, improve the signal-to-noise ratio defined in the ``traditional sense'', this, of course, does not necessarily mean that the transmission performance of the $b$-modulated system would get better. The coupling of solitary modes to the members of the encoded continuous spectrum due to the deviation of the channel from the integrable NLS (e.g. due to the presence of noise, in the simplest scenario) should definitely affect and potentially degrade the performance of the system in hand \cite{pvp19,alb18}, though this question has not been studied in detail so far. From the other side, the possibility to embed solitons in the $b$-modulation renders additional flexibility and new opportunities for the design of such systems with improved characteristics. We restrict ourselves by investigating the effective steadiness of the function $b(k)$ introducing additive white Gaussian noise to the corresponding time-domain profile in the absence and presence of the bound states. However, we note that the signal's propagation itself in the presence of noise may cause additional quality degradation for both continuous and discrete NF spectrum components.} 

\section{Derivation of the properties for $b$-modulated signals}\label{sec:derivation}

\subsection{Direct problem for the Zakharov-Shabat system attributed to $q(t)$ with a finite extent.}\label{subsec:direct}
The forward NFT for the signals $q(t,z)$ evolving according to Eq. (\ref{nlse}), is performed by considering the Zakharov-Shabat system for the two-component function $\varphi(t,k)$ \cite{zs72}, where here and in the following we drop the dependencies of all quantities on $z$ for simplicity: 
\begin{equation}\label{Z-Sh}
\varphi_t +i k \sigma_3 \varphi = Q(t) \varphi, \qquad Q(t) = \begin{pmatrix}
0 & q(t) \\ - \bar q(t) & 0
\end{pmatrix}.
\end{equation}
Here and in the following the overbar \red{($\bar q$)} means the complex conjugate, and  $\sigma_3=\left(\begin{smallmatrix} 1 & 0 \\ 0 & -1
\end{smallmatrix}\right)$.
We assume that for $t\in \mathbb R$, $q\in L^1(\mathbb R)$
and $q(t)=0$ for $|t|>\frac{L}{2}$, for some positive quantity $L>0$, which is our localisation extent.

Let us define the two-component Jost solutions $\Phi^{(j)}(t,k)$ and 
$\Psi^{(j)}(t,k)$, $j=1, \,2$, of 
Eq.~(\ref{Z-Sh}), for $k\in \mathbb R$, fixed by their asymptotic behaviour:
\[
\Phi^{(1)}(t,k) \equiv [\phi_1,\phi_2]^T \to [e^{-ikt},0 ]^T \quad \text{as}\ t\to -\infty, 
\]
\[
\Psi^{(2)}(t,k)\equiv [\psi_1,\psi_2]^T \to [0,e^{ikt}]^T \quad \text{as}\ t\to \infty,
\]
and $\Phi^{(2)} = [-\bar \phi_2 , \bar \phi_1 ]^T$, $\Psi^{(1)} =   [ \bar\psi_2 , -\bar\psi_1]^T$. 
The scattering coefficients $a(k)$ and $b(k)$ 
associated with a given  $q(t)$, are defined through the scattering relation:
\begin{equation}\label{scat-2}
\Phi(t,k) = \Psi(t,k) \begin{pmatrix}
a(k) & -\bar b(k) \\ b(k) & \bar a(k)
\end{pmatrix} \!, \quad k\in \mathbb R,
\end{equation}
with
$$|a(k)|^2 + |b(k)|^2 \equiv 1.$$ In Eq. (\ref{scat-2}) $\Phi=(\Phi^{(1)},\Phi^{(2)})$, and similarly for
$\Psi$. 
Now notice that for finite-extent $q(t)$, $\Phi(t,k)$ and $\Psi(t,k)$ are the entire analytic functions of $k\in\mathbb C$. Moreover, in this case (\ref{scat-2})
holds for all $k\in \mathbb C$ 
with $\bar a$, $\bar b$ replaced by $a^*$, $b^*$,
where the asterisk means the Schwarz reflection:  $\varphi^*(k):=\overline{\varphi(\bar k)}$.
Consequently, 
\begin{equation}\label{det}
a^*(k) a(k) + b^*(k) b(k)\equiv 1,\quad k\in \mathbb C.
\end{equation}

It is well-known that if $q(t)=0$ for $|t|>\frac{L}{2}$ with some $L>0$, 
then the associated spectral functions 
$a(k)$ and $b(k)$ can be expressed via the Fourier transforms of some finitely supported functions
\red{
(see, e.g., \cite{gzl18}, \cite{s04} and Appendix A in \cite{v18-1}).} For the consistency of presentation, we give here a simple proof of this property using the integral representations for the Jost solutions (cf. \cite{s04}).

\begin{thm}\label{th1}
	Let $q\in L^1(\mathbb R)$ be such that $q(t)=0$ for $|t|>\frac{L}{2}$ for some $L>0$.
	Then 
	\begin{itemize}
		\item 
		$\Phi(t,k)= e^{-ikt\sigma_3}$ for $t<-\frac{L}{2}$, and 
		\begin{equation}\label{op-1}
		\Phi(t,k) = e^{-ikt\sigma_3} + \!\int_{-L-t}^t K_1(t,\tau) e^{-ik\tau\sigma_3} d\tau,
		\, \, \, t>-\frac{L}{2};
		\end{equation}
		\item 
		$\Psi(x,k)= e^{-ikt\sigma_3}$ for $t>\frac{L}{2}$, and 
	\end{itemize}
	\begin{equation}\label{op-2}
	\Psi(t,k) = e^{-ikt\sigma_3} + \!\int_{t}^{L-t} K_2(t,\tau) e^{-ik\tau\sigma_3} d\tau,
	\, \, \, t<\frac{L}{2}.
	\end{equation}
	Here $K_j(t,\cdot)\in L^1$, $j=1, \, 2$, are some $2 \times 2$ matrix functions.
\end{thm}

\begin{IEEEproof}[Proof of Theorem \ref{th1}] 
	For any $q(t)\in L^1(\mathbb R)$, the integral representation for $\Phi$  has the form \cite{zs72}:
	\begin{equation}\label{op-1-gen}
	\Phi(t,k) = e^{-ikt\sigma_3} +\int_{-\infty}^t K_1(t,\tau) e^{-ik\tau\sigma_3} d\tau.
	\end{equation}
	Assuming for a moment that $q(t)$ is smooth ($q(t)\in C^1(\mathbb R)$),
	substituting (\ref{op-1-gen}) into (\ref{Z-Sh}), 
	\red{and applying $\int_{-\infty}^\infty (\cdot)e^{ik\tau'} dk$,}
	it follows that  $K_1(t,\tau)$ 
	satisfies the system of equations:
	\begin{equation}\label{goursat-1}
	\begin{aligned}
	& K_1(t,t)-\sigma_3 K_1(t,t) \sigma_3 = Q(t),\\
	& K_{1t}(t,\tau) + \sigma_3 K_{1\tau}(t,\tau) \sigma_3 - Q(t)K_{1}(t,\tau) = 0, \, \, \, \, \tau<t,
	\end{aligned}
	\end{equation}
	where $Q(t)$ is given in Eq.~(\ref{Z-Sh}).
	Decomposing $K_1$ into the diagonal and off-diagonal parts, $K_1^d$ and $K_1^o$, respectively,
	\[
	K_1 = K_1^o + K_1^d,
	\]
	Eq.~(\ref{goursat-1}) then reduces to 
	\begin{equation}\label{goursat-2}
	\begin{aligned}
	& K_1^o(t,t)= \frac{1}{2} Q(t),\\
	& K_{1t}^o(t,\tau)  - K_{1\tau}^o(t,\tau) - Q(t)K_{1}^d(t,\tau) = 0, \quad
	\tau<t, \\
	& K_{1t}^d(t,\tau)  + K_{1\tau}^d(t,\tau) - Q(t)K_{1}^o(t,\tau) = 0, \quad
	\tau<t.
	\end{aligned}
	\end{equation}
	Now changing the variables as $\xi=t+\tau$, $\eta=t-\tau$, and $\tilde K(\xi,\eta):=K_1(t,\tau)$, with $\tilde K_\xi = \frac{1}{2}(K_{1t}+K_{1\tau})$, $\tilde K_\eta = \frac{1}{2}(K_{1t}-K_{1\tau})$, system (\ref{goursat-2}) reduces to the following one:
	\begin{equation}\label{goursat-3}
	\begin{aligned}
	& \tilde K^o(\xi,0)= \frac{1}{2} Q\left(\frac{\xi}{2}\right),\\
	& \tilde K^o_\eta(\xi,\eta) = \frac{1}{2} Q\left(\frac{\xi+\eta}{2}\right)
	\tilde K^d(\xi,\eta), \quad
	\eta>0, \\
	& \tilde K^d_\xi(\xi,\eta) = \frac{1}{2} Q\left(\frac{\xi+\eta}{2}\right)
	\tilde K^o(\xi,\eta), \quad
	\eta>0.
	\end{aligned}
	\end{equation}
	In turn, Eq.~(\ref{goursat-3}) reduces to an integral equation of Volterra type.
	Indeed, integrating (\ref{goursat-3}) we have:
	\begin{equation}\label{int-1}
	\begin{aligned}
	\tilde K^o(\xi,\eta) & = \tilde K^o(\xi,0) + \frac{1}{2} 
	\int_0^\eta Q\left(\frac{\xi+\eta'}{2}\right)\tilde K^d(\xi,\eta')d\eta'\\
	& = \frac{1}{2} Q\left(\frac{\xi}{2}\right) + \frac{1}{2} 
	\int_0^\eta Q\left(\frac{\xi+\eta'}{2}\right)\tilde K^d (\xi,\eta')d\eta', \\
	\tilde K^d(\xi,\eta) & = \frac{1}{2} \int_{-\infty}^\xi 
	Q\left(\frac{\xi'+\eta}{2}\right)
	\tilde K^o(\xi',\eta) d\xi'.
	\end{aligned}
	\end{equation}
	Substituting the second expression from Eq.~(\ref{int-1}) into the first one,
	we arrive at a single integral equation:
	\begin{equation}\label{int-2}
	\begin{aligned}
	&\tilde K^o(\xi,\eta) = \frac{1}{2} Q\left(\frac{\xi}{2}\right) + \frac{1}{4}\int_0^\eta Q\left(\frac{\xi+\eta'}{2}\right)\times \\
	&\times 
	\int_{-\infty}^\xi 
	Q\left(\frac{\xi'+\eta'}{2}\right)
	\tilde K^o(\xi',\eta') \, d\xi' \, d\eta'.
	\end{aligned}
	\end{equation}
	Now notice that for $\xi<-L$, we have $Q\left(\frac{\xi}{2}\right)=0$, and, thus, 
	Eq.~(\ref{int-2}) becomes a homogeneous Volterra integral equation
	(in the domain $\xi<-L$, $\eta>0$), the unique solution of which is 0.
	Therefore, $\tilde K(\xi,\eta)\equiv 0$ for $\xi<-L$, $\eta>0$,
	and, thus, $K_1(t,\tau)=0$ for $t+\tau<-L$. The general case of $q\in L^1$ follows further by approximating $Q$ in (\ref{int-2}) by smooth functions.
	
	Similarly,  $\Psi(t,k)$ has the  representation 
	\[
	\Psi(t,k) = e^{-ikt\sigma_3} +\int_{t}^{\infty} K_2(t,\tau) e^{-ik\tau\sigma_3} d\tau,
	\]
	where,  actually, $K_2(t,\tau)=0$ for $t+\tau >L$, which can be proven by following similar
	arguments as above and taking into account that $Q\left(\xi/2\right)=0$ for $\xi>L$.
\end{IEEEproof}

\begin{cor}\label{cor-1} 
	\begin{enumerate}
		\item 
		In this case, the associated scattering functions $a(k)$ and $b(k)$ have the following integral representations:
		\begin{equation}
		\label{ab}
		a(k) = 1 + \int_0^{2L} \alpha(\tau)e^{ik\tau}d\tau,\quad 
		b(k) =  \int_{-L}^{L} \beta(\tau)e^{ik\tau}d\tau,
		\end{equation}
		with some $\alpha(\tau)\in L^1(0,2L)$, $\beta(\tau)\in L^1(-L,L)$.
		\item
		For $t>L/2$,  $b(k)e^{2ikt}\to 0$ as $k\to\infty$ for $k\in {\mathbb C}_+$,
		and $b^*(k)e^{-2ikt}\to 0$ as $k\to\infty$ for $k\in {\mathbb C}_-$.
		
		Here and below, ${\mathbb C}_\pm = \{k\in {\mathbb C}: \pm \Im k >0 \}$
		\magenta{and $\overline{\mathbb C}_\pm = \{k\in {\mathbb C}: \pm \Im k \ge0 \}$ }.
		\item
		For $t<-L/2$, $b^*(k)e^{-2ikt}\to 0$  as $k\to\infty$ for $k\in {\mathbb C}_+$,
		and $b(k)e^{2ikt}\to 0$ as $k\to\infty$ for $k\in {\mathbb C}_-$.
	\end{enumerate}
\end{cor}

Indeed, 
setting $t=-L/2$ in the scattering relation (\ref{scat-2}), and  using Eq.~(\ref{op-2})
and the fact that $\Phi(-\frac{L}{2},k)=e^{ik\frac{L}{2}\sigma_3}$, it follows 
that $a$ and $b$ have the representations in form of Eq.~(\ref{ab}), where 
$\alpha(\tau) = (K_2)_{22}\left(-\frac{L}{2},\tau-\frac{L}{2}\right)$
and $\beta(\tau) = -(K_2)_{21}\left(-\frac{L}{2},\frac{L}{2}-\tau\right)$.
Here the double subscript $(\cdot)_{ij}$ stands for the corresponding matrix entry.
Items 2) and 3) directly follow from Eq.~(\ref{ab}).

\subsection{Inverse problem attributed to band-limited $b(k)$.}

In the general case $q(t)\in L^1(\mathbb R)$,
we have \cite{ft07}: $a(k) = 1 + \int_0^\infty \alpha(\tau) e^{ik\tau} d\tau$ and 
$b(k) = \int_{-\infty}^\infty \beta(\tau) e^{ik\tau} d\tau$
with $\alpha(\tau)\in L^1(0,\infty)$ and $\beta(\tau)\in L^1(-\infty,\infty)$,
and the set of spectral data 
determining uniquely $q(t)$, is conventionally characterised assuming that $a(k)\ne 0$
for $k\in {\mathbb R}$ and 
all zeros of $a(k)$ in ${\mathbb C}_+$ are  simple;
consequently, the number of these zeros is finite, and $|b(k)|<1$ for all $k\in \mathbb R$.
\magenta{With these assumptions,} the characteristic spectral data consist of $b(k)$, $k\in \mathbb R$
and the discrete set $\{k_j, c_j\}_1^N$ (\red{given, in general,  independently of $b(k)$}; particularly, it
can be empty), 
where $k_j$ with $\Im k_j>0$, $j=1,\dots,N$,
are the zeros of $a(k)$,
and $\{c_j\}_1^N$ are the associated norming constants. Moreover, the inverse mapping can be described as follows \cite{ft07}:
\begin{enumerate}
	\item 
	Given $b(k)$ and $\{k_j\}_1^N$, construct $a(k)$ in accordance with  (\ref{det})
	for $k\in\mathbb R$:
	\begin{equation}\label{a-kj}
	a(k) = \prod_{j=1}^N \frac{k-k_j}{{k- \bar k_j}}\exp\left\{\frac{1}{2\pi i}\int_{\mathbb R}
	\frac{\log(1-|b(s)|^2)}{s-k}ds \right\};
	\end{equation}
	\item 
	Define the reflection coefficient 
	\begin{equation}\label{r}
	r(k):=b(k)/a(k), \qquad k\in \mathbb R;
	\end{equation}
	\item
	Solve the Riemann--Hilbert problem (RHP): find a $2\times 2 $ function $M(t,k)$ satisfying the 
	following conditions:
	\begin{itemize}
		\item 
		As a function of $k$, $M$ is meromorphic in ${\mathbb C}\setminus
		{\mathbb R}$.
		\item
		The limiting values $M_\pm(t,k)$, $k\in \mathbb R$  of $M(t,k)$ as $k$ approaches
		the real line from ${\mathbb C}_\pm$ are related by 
		\begin{equation}\label{jump-1}
		M_+(t,k)=M_-(t,k)J(t,k), \qquad k\in \mathbb R,
		\end{equation}
		where 
		\begin{equation}\label{J-1}
		J(t,k)=\begin{pmatrix}
		1+|r(k)|^2 & r^*(k) e^{-2ikt} \\  r(k) e^{2ikt} & 1
		\end{pmatrix}.
		\end{equation}
		\item
		$M(t,k)\to I$ as $k\to\infty$.
		\item 
		The singularities of $M$ are characterised as follows:
		$M^{(1)}(t,k)$ has simple poles at $\{k_j\}_1^N$ and 
		$M^{(2)}(t,k)$ has simple poles at $\{\bar k_j\}_1^N$ such that the following 
		residue conditions hold:
		\begin{equation}\label{res}
		\Res  M^{(1)}(t,k)\Big|_{k=k_j} = c_j e^{2ik_j t}M^{(2)}(t,k_j),
		\end{equation}
		\begin{equation}\label{res-2}
		\Res  M^{(2)}(t,k)\Big|_{k=\bar k_j} = -\bar c_j
		e^{-2i \bar k_j t} M^{(1)}(t,\bar k_j).
		\end{equation}
		\red{Here,   $M^{(j)}$, $j=1, \, 2$, denotes the $j$-th column of a $2\times 2$ matrix $M$: $M=(M^{(1)},M^{(2)})$.}
	\end{itemize}
	\item
	Having the RHP solved, $q(t)$ can be obtained by
	\red{
\begin{equation}\label{q-M}
	q(t)= 2i \overset{[1]}{M}_{12}(t),
		\end{equation}
	where the subscript ``$12$'' denotes the corresponding matrix entry and 
	the matrix $\overset{[1]}{M}(t)$ emerges from the large-$k$ development of $M(t,k)$:
\begin{equation}\label{asympt}
	M(t,k) = I + \frac{\overset{[1]}{M}(t)}{k} + O(k^{-2}), \qquad k\to\infty,
\end{equation}
	where $I$ is the identity matrix.
	}
\end{enumerate}
\begin{rem}
\red{In the framework of the direct scattering problem (given $q(t)$, determine the scattering data)},
	$M(t,k)$ is related to Jost solutions of the Zakharov-Shabat problem (\ref{Z-Sh}) as follows
	\[
	M(t,k)=\begin{cases}
	\left(\frac{\Phi^{(1)}(t,k)}{a(k)}, \Psi^{(2)}(t,k) \right)e^{ikt\sigma_3}, 
	& k\in {\mathbb C}_+,\\
	\left(\Psi^{(1)}(t,k), \frac{\Phi^{(2)}(t,k)}{a^*(k)}\right)e^{ikt\sigma_3},
	& k\in {\mathbb C}_-.
	\end{cases}
	\]
\end{rem}

\red{
Our main result consists in the characterization of signals $q(t)$
having the $b$-coefficients in the form of the Fourier transform 
of a function with limited (bounded) support.
}
\red{
Let a function $b(k)\not\equiv 0$, $k\in\mathbb{C}$, be given such that
\begin{enumerate}
		\item[(i)] $b(k) = \int_{-L}^{L}\beta(\tau)e^{ik\tau}d\tau$
		with some $\beta(\tau)\in L^1(-L,L)$;
		\item[(ii)] The function $G(k):=1-b^*(k)b(k)$ has no zeros for 
		$k\in \mathbb R$ (or, equivalently, $G(k)>0$ for $k\in {\mathbb R}$). 
	\end{enumerate}}
\begin{definition}\label{f_b}
\red{Define ${\cal F}_b$ as the set of all $q\in L^1$ such that 
the spectral functions $b(k)=b(k;q)$ and $a(k)=a(k;q)$ associated to $q$ by the direct mapping through (\ref{Z-Sh}) and (\ref{scat-2}), satisfy the following conditions:}
\begin{enumerate}
		\item[(a)] \magenta{ $b(k;q)$ coincides with the prescribed function $b(k)$ satisfying conditions (i) and (ii) above (particularly, we assume that $a(k;q)$ is not zero for real $k$);}
		\item[(b)]\red{ All zeros of  $a(k;q)$ for $k\in{\mathbb{C}_+}$ are simple.}
	\end{enumerate}
\end{definition}
\red{\begin{notation}\label{a_b}
Denote by ${\cal A}_b$ the set of all zeros of 
	$G(k)$.
\end{notation}}
\begin{thm}\label{main}
	\begin{enumerate}
			\item \red{The set ${\cal F}_b$ consists of infinitely many elements;}
		\item \red{
		For any $q\in {\cal F}_b$, $q(t)=0$  for $|t|>L/2$; }
		\item 
		\magenta{
		Each particular $q\in{\cal F}_b$ is uniquely specified by a finite subset 
		$\{k_j\}_{1}^N$ (including the empty set) of 
		${\cal A}_b\cap {\mathbb C}_+$. This subset constitutes the set of simple zeros 
		of the spectral function $a(k)$ in $\mathbb C_+$ associated to this $q$,
		which can be expressed for $k\in \overline{\mathbb C}_+$  by (\ref{a-kj})  in terms of $b(k)$, $k\in\mathbb R$ and $\{k_j\}_{1}^N$.}
	\end{enumerate}
\end{thm}
\red{
\begin{rem}
In the case $b(k)\equiv 0$, it follows from (\ref{det}) that the associated $a(k)$ 
has no zeros and thus, by (\ref{a-kj}),  $a(k)\equiv 1$; consequently, $q(t)$ with such spectral data is the trivial one:
$q(t)\equiv 0$.
\end{rem}
}

\red{
\noindent \textit{Proof of Theorem \ref{main}}.
The proof is based on using the 
flexibility of the RHP formalism: (i) the same $q(t)$ can be retrieved from the solutions of different RHPs; 
(ii) we can proceed from one RHP to another (that produces the same $q(t)$) by 
appropriately ``transforming'' the original RHP, 
e.g., factorizing the jump matrix and absorbing the factors into the solution of a new RHP
problem, in specific domains of the complex plane.
}

First, notice that in our case $b(k)$ is analytic in $\mathbb C$ and thus the norming constants
are determined by $b(k)$ and $a(k)$: 
\begin{equation}\label{norm-c} c_j=\frac{b(k_j)}{ \dot a(k_j)},
\end{equation}
where the overdot \red{($\dot a$)} means the derivative with respect to $k$.
\begin{IEEEproof}[\red{Proof that ${\cal F}_b$ consists of infinitely many elements}]
\red{
We notice that  $b(k)$ and $G(k)$ are the entire functions
of order at most 1 \cite{levin}. This follows from the following estimates:
\[
|b(k)|=\left|\int_{-L}^L \beta(\tau)e^{ik\tau}d\tau \right|\le  e^{L|k|} 
\int_{-L}^L |\beta(\tau)|d\tau,
\]
and
\[
\begin{aligned}
|G(k)|&\le 1+\left|\int_{-L}^{L} \int_{-L}^{L} e^{i k (\xi+\eta)}\beta(\xi)
\overline{\beta(-\eta)}\,d\xi\,d\eta\right|\le\\
&\le 1+e^{2L|k|} \int_{-2L}^{2L}\left|B(\tau)\right| d\tau,
\end{aligned}
\]
with $$B(\tau) =\int_{-L}^{L}\beta(\eta)\overline{\beta(\eta-\tau)}d\eta 
\in L^1(-2L,2L).$$
Actually, $b(k)$ is the entire function of order 1 and type $L$ provided
$\beta(\tau)$ does not vanish almost everywhere in any neighborhood of $L$ or $-L$,
see \cite[subsec.~6.9.1]{boas}; similarly for $G(k)$.
}
\red{
Next, for an entire function of order at most 1, 
 the Hadamard factorisation theorem, see \cite[p. 26]{levin}, 
implies  that $G(k)$ can be written as a product:
\begin{subequations}\label{prod}
\begin{equation}\label{prod-1}
G(k) = k^m e^{p_1 k+p_2}\prod_{n=1}^{M}\left(1-\frac{k}{k_n}\right)e^{\frac{k}{k_n}},
\end{equation}
or 
\begin{equation}\label{prod-0}
G(k) = k^m e^{p_1 k+p_2}\prod_{n=1}^{M}\left(1-\frac{k}{k_n}\right),
\end{equation}
\end{subequations}
where $m\ge 0$, $p_j\in \mathbb{C}$, and $\{k_n\}_1^M$ with $M\le\infty$ are the zeros of $G(k)$.
}
\red{
Assume that $M<\infty$. Then it follows from (\ref{prod}) that 
\begin{equation}\label{poly}
G(k)=e^{\tilde p_1 k+p_2}P(k),
\end{equation}
 with some $\tilde p_1$ and $p_2$, where 
$P(k)$ is a polynomial of degree $m+M$.
On the other hand, we notice that for $k=x\in \mathbb R$, 
$b(x)=\int_{-L}^{L}\beta(\tau) e^{ix\tau} d\tau \to 0$ as $x\to\pm \infty$
and thus $G(x)\to 1$.
Combining this with $G(x)=e^{\tilde p_1 x+p_2}P(x)$ evaluated as $x\to +\infty$,
and as $x\to -\infty$, we conclude that $\tilde p_1$ must be $0$ and thus $G(x)\equiv 1$,
which is in contradiction  with $b(k)\not\equiv 0$.
}
\end{IEEEproof}

It follows from (\ref{det}) that the set ${\cal A}_b$ (determined by $b(k)$
and symmetric w.r.t. the real axis)
is a union of zeros of $a(k;q)$ and $a^*(k;q)$. Consequently, all zeros of 
$a(k;q)$ in ${\mathbb C}_+$ \red{associated with $q\in {\cal F}_b$} (the eigenvalues of (\ref{Z-Sh}))  must be contained in 
${\cal A}_b\cap {\mathbb C}_+$. 

Let us choose any finite (particularly, it can be empty)
subset $\{k_j\}_1^N$
from ${\cal A}_b\cap {\mathbb C}_+$, construct $a(k)$
\magenta{ for $k\in \overline{\mathbb C}_+$ by 
(\ref{a-kj}) (accordingly, $a^*(k)$ is determined for $k\in \overline{\mathbb C}_-$), and determine $r(k)$ for $k\in \bar{\mathbb C}_+$ by $r(k)=b(k)/a(k)$ (cf. (\ref{r}))
as well as $r^*(k)$ for $k\in \overline{\mathbb C}_-$ by $r^*(k)=b^*(k)/a^*(k)$. }
Our main point is that using $b(k)$, $a(k)$, and $\{k_j,c_j\}_1^N$,
specified above, as the spectral data and the input to RHP, Eqs.~(\ref{jump-1})--(\ref{res-2}), one always arrives at such $q(t)$ that 
$q(t)=0$ for $|t|>L/2$.

\begin{IEEEproof}[Proof that $q(t)=0$ for $t>L/2$] The proof is based 
on the \red{transformation of RHP (\ref{jump-1})--(\ref{res-2}): $M\mapsto \hat M$, }suggested by the following algebraic 
	factorization of $J$ in (\ref{J-1}):
	\begin{equation}\label{fact-1}
	J(t,k) = \begin{pmatrix}
	1 &  r^*(k) e^{-2ikt} \\  0 & 1
	\end{pmatrix}
	\begin{pmatrix}
	1 & 0 \\  r(k) e^{2ikt} & 1
	\end{pmatrix}, \qquad k\in \mathbb R.
	\end{equation}
	
Recall that the RHP (\ref{jump-1})--(\ref{res-2}) (particularly, the residue
conditions (\ref{res}), (\ref{res-2})) has been formulated under assumption that 
$a(k)$ has no real zeros; otherwise (\ref{res}), (\ref{res-2}) are not correct.

\red{For all $t>L/2$, define  $\hat M(t,k)$ by}
	\begin{equation}\label{hat-m-1}
	\hat M(t,k):= \begin{cases}
	M(t,k) \begin{pmatrix}
	1 & 0 \\  -r(k) e^{2ikt} & 1
	\end{pmatrix}, & k\in {\mathbb C}_+,\\
	M(t,k) \begin{pmatrix}
	1 &  r^*(k) e^{-2ikt} \\  0 & 1
	\end{pmatrix}, & k\in {\mathbb C}_-.
	\end{cases}
	\end{equation}
	Notice that the triangular matrix factors in (\ref{hat-m-1})  considered 
	(for any fixed $t>L/2$) as functions of $k$  in the corresponding half-planes
	are such that (i)  they are meromorphic in the respective half-planes
	and (ii) they approach $I$ \red{exponentially fast} as $k\to\infty$. The latter follows from 
	Corollary \ref{cor-1}, item 2,
	and the fact that $a(k)\to 1$ as $k\to\infty$ for $k\in {\mathbb C}_+$.
	\red{
	Particularly, this implies that the large-$k$ 
	expansion for $\hat M(t,k)$ coincides  with that for $M(t,k)$, see (\ref{asympt}),
	and thus, determining $\hat q(t)$ from $\hat M(t,k)$ in the same way as 
	$q(t)$ is determined from $M(t,k)$, see  (\ref{q-M}),
	}
 we have:
	\begin{equation}\label{qq-1}
	\hat q(t) = q(t), \quad t>L/2.
	\end{equation}

	On the other hand, $\hat M(t,k)$ can be characterized as the solution 
	of the RHP with the trivial jump conditions: \red{find $\hat M(t,k)$  analytic
	in ${\mathbb C}\setminus{\mathbb R}$ and satisfying the properties
	\begin{align}\label{rhp-mod}
	\hat M_+(t,k) &= \hat M_-(t,k) , \qquad k\in \mathbb R,\nonumber\\
	\hat M(t,k) & \to I, \qquad k\to \infty.
	\end{align}
	Indeed, in view  of the factorization (\ref{fact-1}), 
	the jump conditions across $\mathbb R$ for $\hat M$ turn out to be trivial, 
	}
and the statement is obvious in the case when $a(k)$ has no zeros in ${\mathbb C}_+$.
	If $a(k_j)=0$ for some $k_j\in {\mathbb C}_+$, we
	evaluate $\hat  M^{(1)}(t,k)$ 	as $k\to k_j$ by using   
	\[
	\hat  M^{(1)}(t,k) = M^{(1)}(t,k) - \frac{b(k)}{a(k)} e^{2ik t}M^{(2)}(t,k)
	\]
	that follows from Eq.~(\ref{hat-m-1}).
	Taking into account (\ref{res}), it then follows that, as $k\to k_j$,
	\begin{equation}\nonumber
	\begin{aligned}
	&\hat  M^{(1)}(t,k) = \frac{1}{k-k_j}\frac{b(k_j)}{\dot a(k_j)} e^{2ik_j t}M^{(2)}(t,k_j) 
	+O(1) \\ &- \left(\frac{b(k_j)}{\dot a(k_j)(k-k_j)}e^{2ik_j t}M^{(2)}(t,k_j) + O(1)\right) 
	= O(1).
	\end{aligned}
	\end{equation}
	Therefore, $\hat  M^{(1)}(t,k)$ has no singularity at $k =k_j$.
	Similarly for $\hat  M^{(2)}(t,k)$ at $k =\bar k_j$.
		\red{
The trivial jump conditions across $\mathbb R$ in  (\ref{rhp-mod}) 
imply that $\hat M(t,k)$ is, in fact, analytic in the whole complex plane.
Also it is
bounded at infinity; moreover, it approaches the identity matrix as $k\to\infty$.
Then, by the Liouville theorem, $\hat M(t,k)\equiv I$ and thus, in view of 
(\ref{q-M}) and (\ref{asympt}), $\hat q(t)\equiv 0$. Finally, in view of 
 Eq.~(\ref{qq-1}),
} we have
	$q(t)= 0$ for $t>L/2$.
\end{IEEEproof}
\begin{IEEEproof}[Proof that 
	$q(t)=0$ for $t<-L/2$]
	Like above, the proof is based on 
	the deformations of the (original) RHP,
	Eqs.~(\ref{jump-1})--(\ref{res-2}).
	But now it is convenient to do the appropriate transformation in two steps:
	\red{$M\mapsto \tilde M \mapsto \check M$}.
	
	Step 1: \red{$M\mapsto \tilde M$.} Define 
	\begin{equation}\label{tilde-m}
	\tilde M(t,k):= \begin{cases}
	M(t,k) \begin{pmatrix}
	a(k) & 0 \\  0 & \frac{1}{a(k)}
	\end{pmatrix}, & k\in {\mathbb C}_+,\\
	M(t,k) \begin{pmatrix}
	\frac{1}{a^*(k)} &  0 \\  0 & a^*(k)
	\end{pmatrix}, & k\in {\mathbb C}_-.
	\end{cases}
	\end{equation}
	Then, it follows from Eqs.~(\ref{jump-1}) and (\ref{J-1}) that 
	$\tilde M(t,k)$ satisfies the following jump conditions for $k\in \mathbb R$:
	$\tilde M_+(t,k) = \tilde M_-(t,k)\tilde J(t,k)$, where
	\begin{equation}\label{tilde-j}
	\begin{aligned}
	\tilde J(t,k) & = \begin{pmatrix}
	a^*(k)	 &  0 \\  0 & \frac{1}{a^*(k)} 
	\end{pmatrix}
	\begin{pmatrix}
	1+|r(k)|^2 &  r^*(k) e^{-2ikt} \\  r(k) e^{2ikt} & 1
	\end{pmatrix}\\
	&\times
	\begin{pmatrix}
	a(k) & 0 \\  0 & \frac{1}{a(k)}
	\end{pmatrix} \\
	& =
	\begin{pmatrix}
	1	 &  0 \\  \frac{b(k)}{a^*(k)}e^{2ikt} & 1
	\end{pmatrix}
	\begin{pmatrix}
	1	 &  \frac{b^*(k)}{a(k)}e^{-2ikt} \\ 0  & 1
	\end{pmatrix}.
	\end{aligned}
	\end{equation}
	
Step 2: \red{$\tilde M\mapsto \check M$}. The triangular factorization in (\ref{tilde-j}) suggests introducing the second
	RHP deformation step, defining $\check M$ by
	\begin{equation}\label{hat-m-2}
	\check M(t,k):= \begin{cases}
	\tilde M(t,k) \begin{pmatrix}
	1 &  -\frac{b^*(k)}{a(k)} e^{-2ikt} \\  0 & 1
	\end{pmatrix}, & k\in {\mathbb C}_+ ,\\
	\tilde M(t,k)\begin{pmatrix}
	1 & 0 \\  \frac{b(k)}{a^*(k)} e^{2ikt} & 1
	\end{pmatrix}, & k\in {\mathbb C}_-.
	\end{cases}
	\end{equation}
	Now notice that the triangular factors in (\ref{hat-m-2})
	are again meromorphic in the respective half-planes and, in view of 
	Corollary \ref{cor-1}, item 3, they approach $I$ exponentially fast as $k\to\infty$.
	Consequently, for $\check q(t)$ obtained from the large-$k$ 
	asymptotics of $\check M(t,k)$ \red{by (\ref{q-M}) and (\ref{asympt})} we have:
	\begin{equation}\label{qq-2}
	\check q(t) = q(t), \quad t<-L/2.
	\end{equation}

	On the other hand, reasoning as in the case $t>L/2$, we can show that $\check M(t,k)$ 	has no singularities in $\mathbb C\setminus \mathbb R$ and thus $\check M(t,k)$ can be characterised as the solution of the (piecewise analytic) RHP \red{(\ref{rhp-mod}) }with trivial jump conditions. \red{ As above, this
	implies that $\check M(t,k)\equiv I$
	and thus  $\check q(t)\equiv 0$.} In view of Eq.~(\ref{qq-2}), 
	$q(t)= 0$ for $t<-L/2$, which completes the proof of Theorem \ref{main}.
\end{IEEEproof}

\section{Generation of localised b-modulated profiles containing solitons}

According to the discussion above, in order to embed the discrete spectrum components \red{into} the signal generated via the $b$-modulation \red{method}, which would not destroy the localisation of the signal in the time domain, the embedded discrete eigenvalues, $k_{\text{eig}}\in\{k_j\}_1^N$, must satisfy the condition, following from Eq.~(\ref{det}) and item 3 of Theorem \ref{main}: 
\begin{equation}
b^*(k_{\text{eig}}) b(k_{\text{eig}})=1. \label{cond}
\end{equation}
For the known expression for $b(k)$, as it occurs in the optical transmission tasks employing \red{the} $b$-modulation, we can numerically seek for such points in the upper complex half-plane of parameter $k$. Then, these points give us the location where we can place our solitary modes without destroying the complete localisation of the time domain signal.

In the $b$-modulation approach, the signal power is manipulated by scaling of the modulated waveforms. However, this adjustment is typically performed numerically because of the non-trivial dependency between $b(k)$ and $q(t)$ scalings. In particular, let $b(k)=A u(k)$, where $u(k)$ is the waveform modulated in a known way independently of the desired signal power, and assume that we do not have a discrete spectrum. The signal energy, given by the expression through the nonlinear spectrum functions as \cite{gzl18} 
$$\epsilon=-1/\pi\int_{-\infty}^{\infty}\log\left(1-A^2| u(k)|^2\right)dk,$$ 
together with the desired time support value $L$, define the average signal power $P=\epsilon/L$ (in normalised units). Thus, having defined the particular signal power and modulation type, we can calculate the scaling factor $A$ and, therefore, further define the locus of ``allowed'' eigenvalues, i.e. the eigenvalues that would not destroy the exact localisation, exploiting the theoretical results from Sec. \ref{sec:derivation}.

\begin{figure}[tbh]
	\centering
	\includegraphics[width=.45\textwidth]{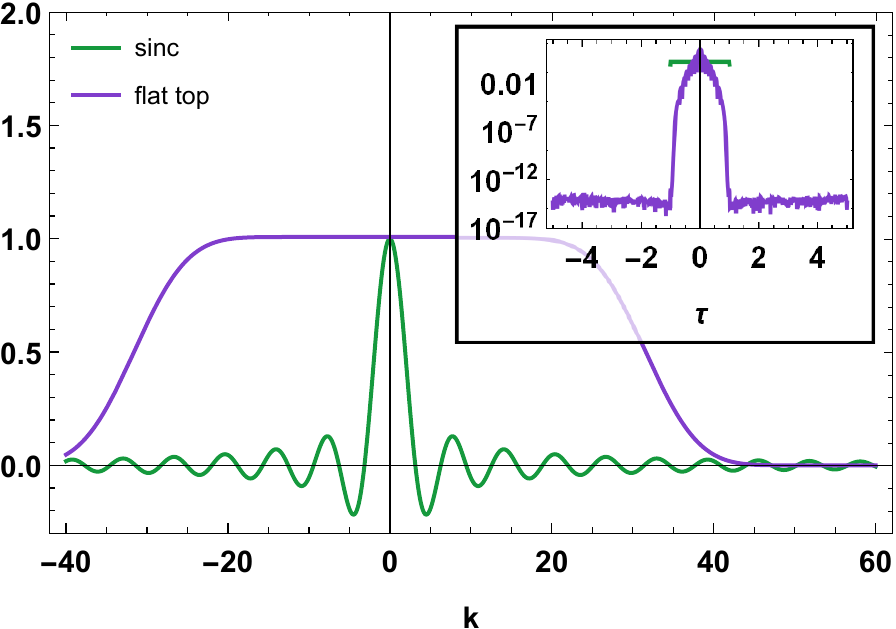}
	\caption{The waveforms used in our work as an example for the illustration of $b$-modulation method with their Fourier transforms in the inset.\label{fig:waveforms} }
\end{figure}

Here we present the procedure for \red{the} generation of a $b$-modulated signal with \red{embedded} discrete NF eigenmodes, which conserves the signal localisation, employing two simple carrier waveforms that are commonly used within the $b$-modulation approach \cite{w17,gzl18}: the Nyquist shape, i.e. the sinc function in the $k$-domain and rectangular profile in the corresponding Fourier-conjugated domain, and the flat-top window carrier function, introduced in \cite{gzl18} as a way of overcoming the $b$-modulated signal power constraint. The studied waveforms used for the $b(k)$ modulation with their corresponding Fourier images are given in Fig.~\ref{fig:waveforms}.

\begin{figure}[t!]
	\centering
	\includegraphics[width=.45\textwidth]{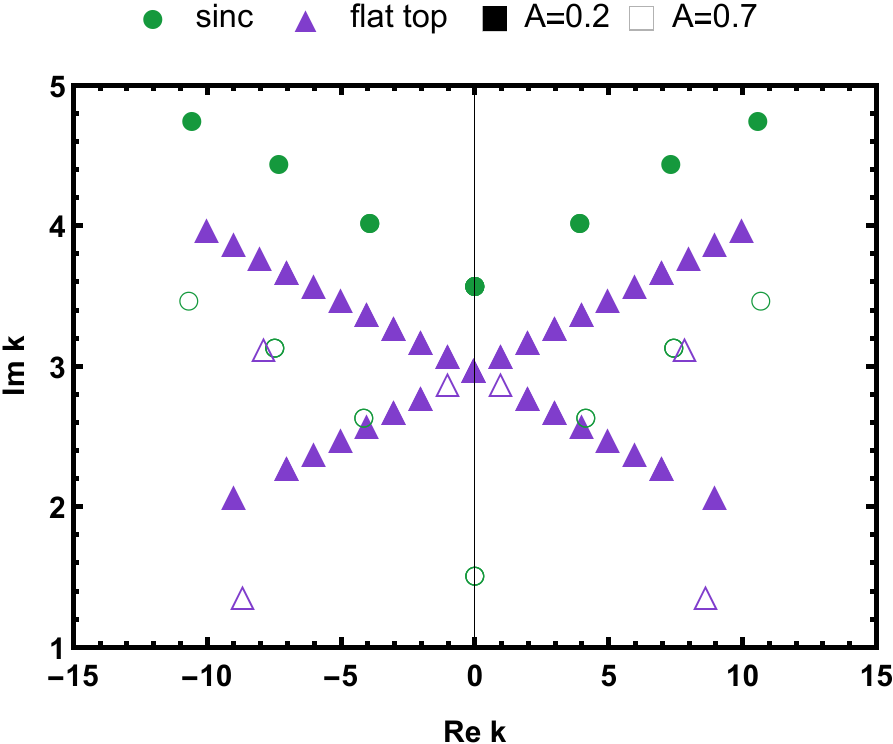}
	\caption{\red{The points 
	from ${\cal A}_b$ 
	 available for  placing the eigenvalues at, when keeping the exact localisation. Circles and triangles distinguish different $b$-shapes, while their filling (or its absence) identifies the level of the scaling factor $A$. As we see, the points depend significantly on the value of $A$ used for signal power manipulation. \label{fig:eigs} }}
\end{figure}

Depending on the value of the scaling factor $A$, these functions have points in the complex plane of $k$, which can be used to implant our eigenvalues at, while keeping the exact localisation of the resulting $q(t)$ profile. Of course, the numerical search cannot guarantee that we have identified all appropriate points, but at the moment we just need to find some of them to illustrate the idea. Typically, for communication purposes we do not use eigenvalues with large real and/or imaginary parts. This occurs in view of the numerical issues associated with the inverse NFT computation for high-amplitude solitons,  or since, e.g., a nonlinear eigenmode with a large real part of its $k_\text{eig}$ would rapidly escape from the dedicated time-window during the signal propagation. The numerically found set of points, which can be used as an eigenvalue locus for our studied $b(k)$ waveforms and different scaling factors, are given in Fig.~\ref{fig:eigs}. Note that according to Theorem \ref{main}, for any band-limited $b(k)$ we always have an infinite number of such points.

\begin{figure}[ht]
	\centering
	\begin{tabular}{c}
		\includegraphics[width=.45\textwidth]{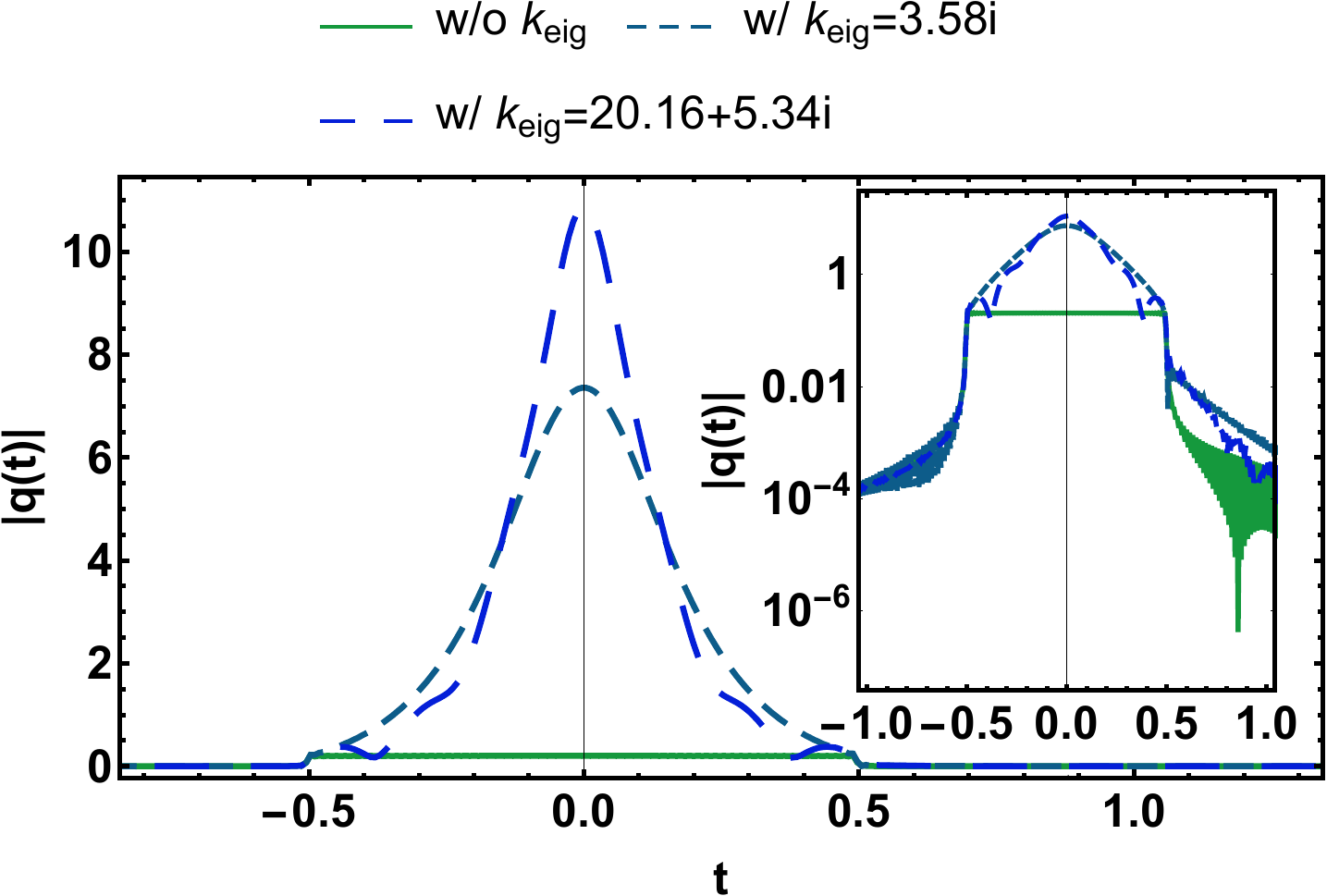}\\
		(a)\\
		\includegraphics[width=.45\textwidth]{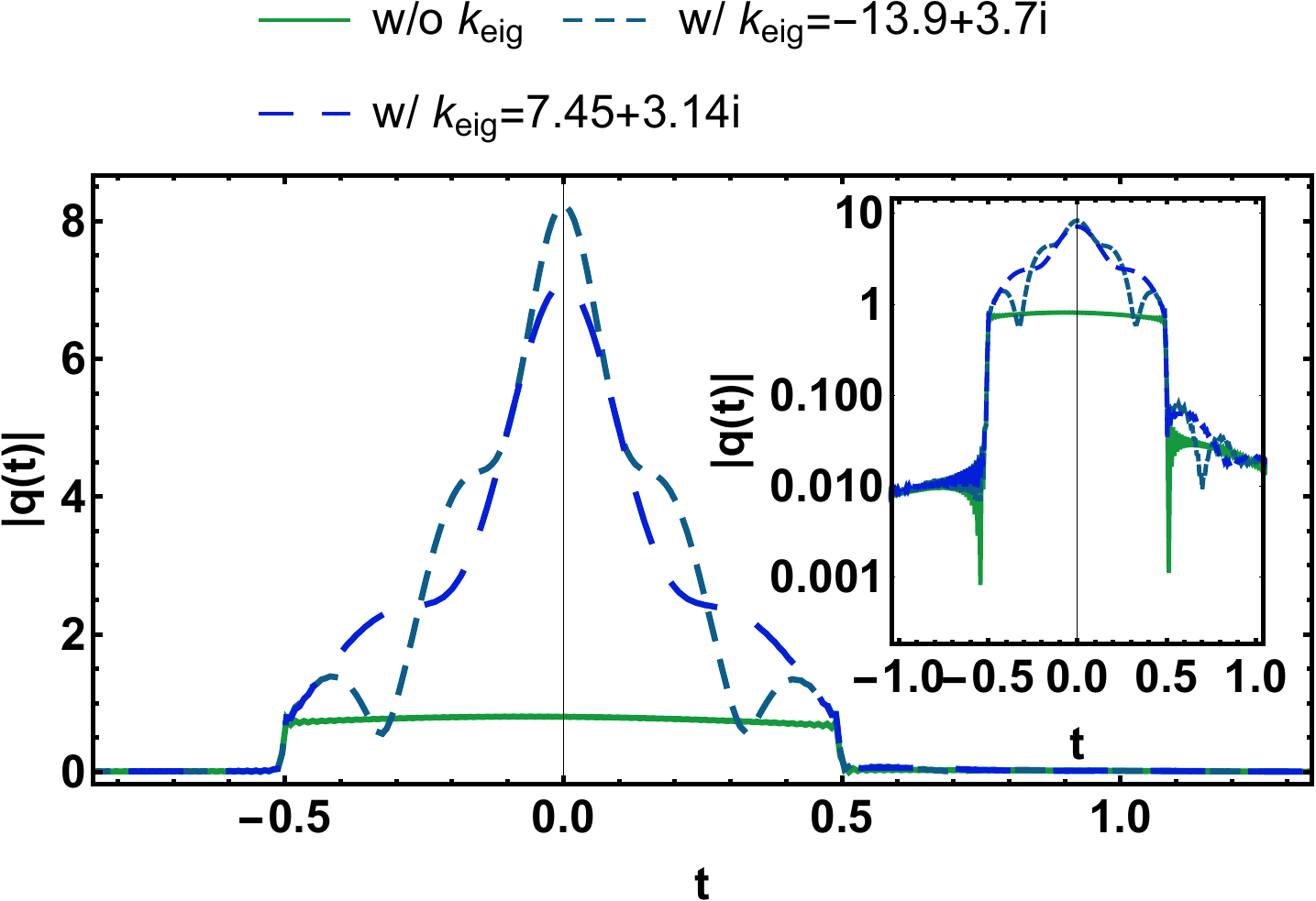}\\
		(b)
	\end{tabular}
	\caption{The signals, generated from the Nyquist waveform via INFT with (dashed) and without (solid) additional eigenvalues embedded, for scaling factors (a) $A=0.2$ and (b) $A=0.7$, and different eigenvalues $k_{\text{eig}}$, marked in the figure.\label{fig:sincs} }
\end{figure}

\begin{figure}[ht]
	\centering
	\begin{tabular}{c}
		\includegraphics[width=.45\textwidth]{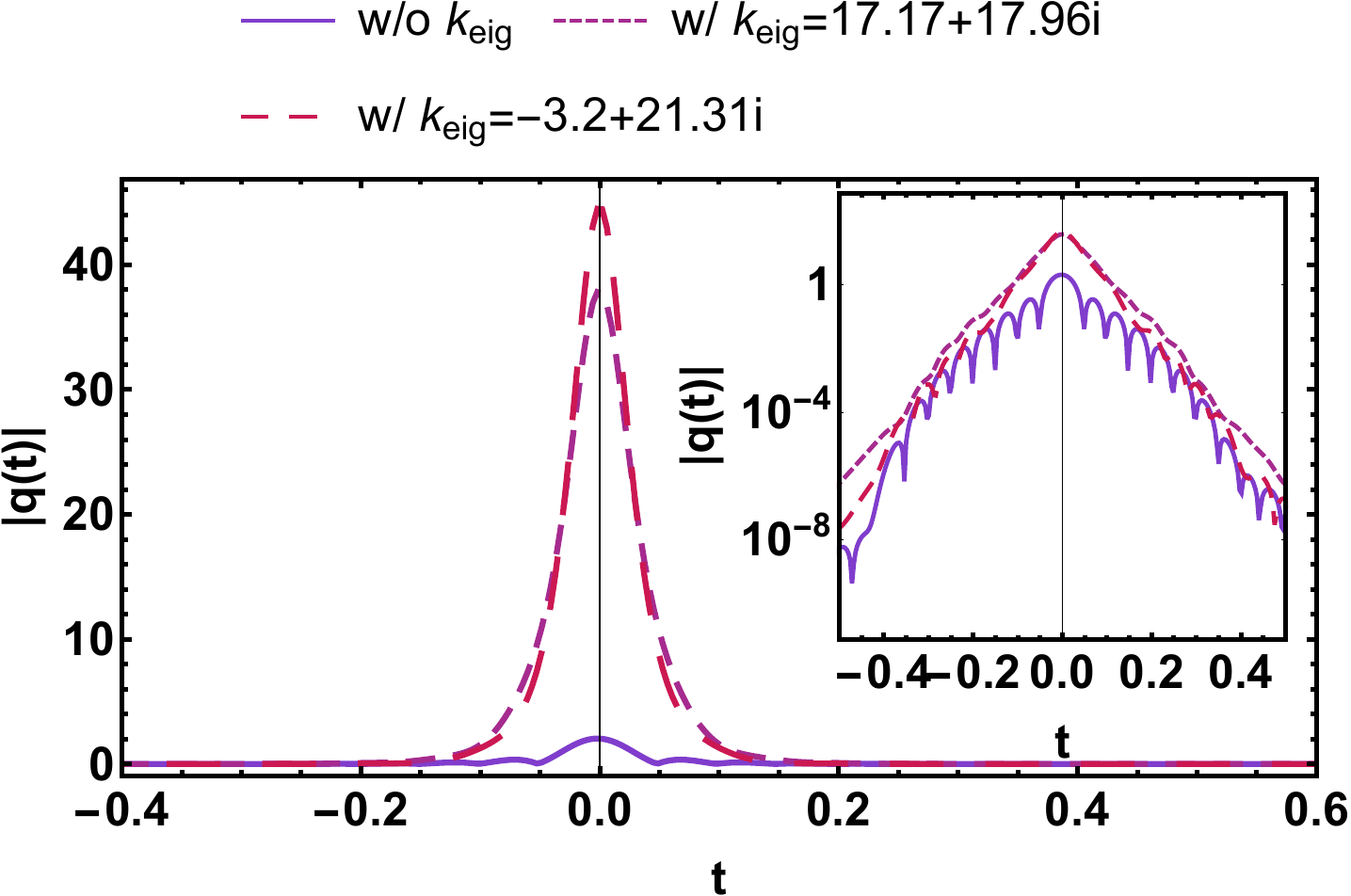}\\
		(a)\\
		\includegraphics[width=.45\textwidth]{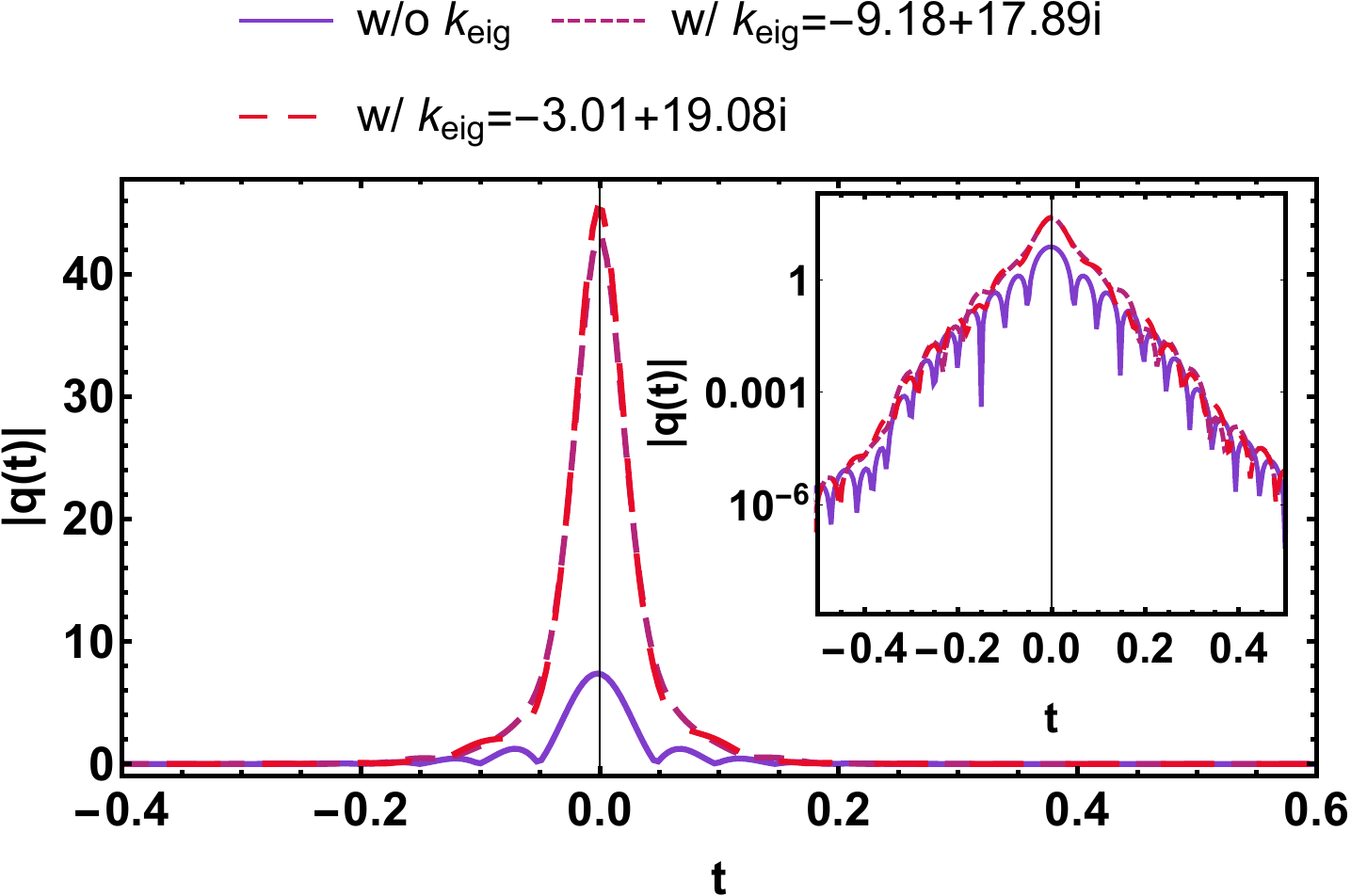}\\
		(b)
	\end{tabular}
	\caption{The signals, generated from the flat-top waveform via INFT with (dashed) and without (solid) additional eigenvalues embedded, for scaling factors (a) $A=0.2$ and (b) $A=0.7$, and different eigenvalues $k_{\text{eig}}$, marked in the figure.\label{fig:flats} }
\end{figure}

The procedure of adding the eigenmodes to the $b$-modulated signal while keeping its exact localisation, is as follows.
\begin{enumerate}
	\item[(i)] Modulate the waveform $u(k)$ with the given information and according to the desired temporal support $L$ of the signal.
	\item[(ii)] Choose the desired signal power (without eigenvalues, as in \cite{gzl18}) and find the appropriate scaling factor $A$. Further define the $b$-function as $b(k)=Au(k)$.
	\item[(iii)] For this $b(k)$, find point(s) $k_{\text{eig}}\in{\mathbb C}_+$, which satisfy $b^*(k_{\text{eig}}) b(k_{\text{eig}})=1$.
	\item[(iv)] Derive corresponding $a(k)$ via Eq.~(\ref{a-kj}), and calculate the resulting $r(k)$ via Eq.~(\ref{r});
	\item[(v)] For each eigenvalue, calculate the corresponding $b_{\text{eig}}:=b(k_{\text{eig}})$, which uniquely defines the respective norming constant $c_\text{eig}$ via Eq.~(\ref{norm-c}).
	\item[(vi)] Generate the signal from the scattering data $r(k)$ and set of $\{k_{\text{eig}}, c_{\text{eig}}\}$ via any inverse NFT procedure \cite{tpl17}, e.g. through the Darboux transform \cite{lab17,a16} or by solving the inverse problem directly with the account of discrete modes. 
\end{enumerate}
We perform the numerical mapping to the time domain from the scattering data using the layer-peeling algorithm (in particular, its fast implementation \cite{wv16,wv17}) with the subsequent Darboux transform \cite{lab17,a16} that adds discrete nonlinear modes to the continuous ones without affecting the latter. The whole procedure follows the scheme given above. The results of the signal generation for both initial waveforms used for $b$-modulation and different scaling factor $A$ values are given in Figs.~\ref{fig:sincs}--\ref{fig:flats}.

We observe that the signals with additional solitonic eigenvalues have at least not \red{worse} localisation than the initial $b$-modulated signal without discrete eigenmodes, in accordance with our theory. However, the numerical algorithms introduce additional errors, which somewhat deteriorates the expected perfect localisation of the resulting $q(t)$ profile. It can be  better seen from the logarithmically scaled plots, given in the insets, that the signal tails decay rates for the profiles with and without additional discrete eigenmodes  coincide almost exactly. In spite of the observed insignificant numerical errors, the results in Figs.~\ref{fig:sincs}--\ref{fig:flats} evidently confirm the correctness of the analytical statements presented in our work. \red{While it is sufficient to compute the signal only within the waveform extent region, 
we intentionally left some (almost) zero ``wings'' to the right and to the left of each resulting localised pulse 
in order to better 
 visualise the localisation.}

\red{
\section{Possible sources of the transmission degradation}
The addition of the eigenvalues to the $b$-modulated signal would be beneficial if this process provides some additional perspectives for the modulation or improvement of the transmission quality. Nonetheless, this process may also introduce some additional penalties because of the complex structure of the transmission line.}

\begin{figure}[h]
\begin{center}
\includegraphics[width=.45\textwidth]{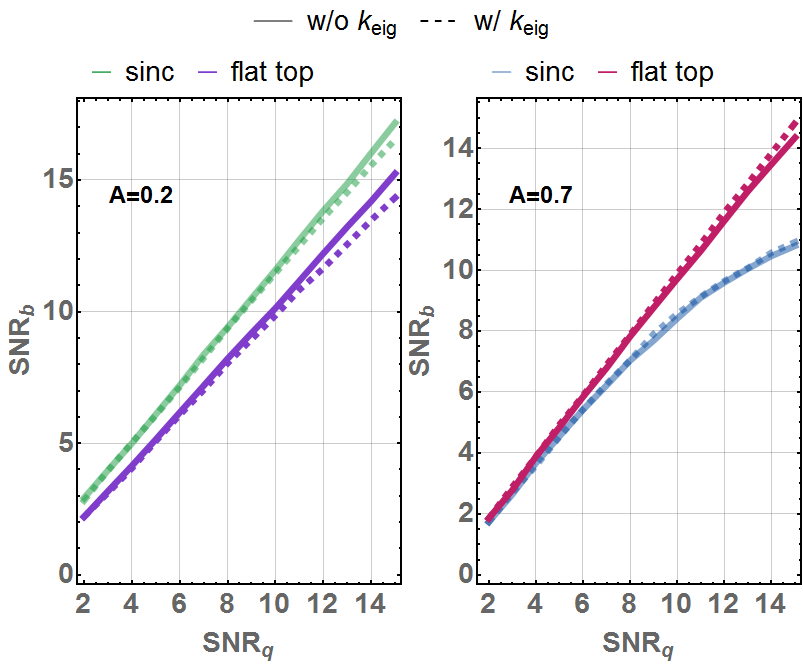}
\end{center}
	\caption{\red{The effective SNR of the $b(k)$ waveform, arising from the noise in the time domain, characterised by the SNR of the $q(t)$ function, for different values of the waveform amplitude.\label{fig:snrs} }}
\end{figure}

\red{It is known that the deviations of the optical channel from the integrable NLS  lead to the effective interaction between continuous and discrete NF spectra \cite{pvp19, ga19}, whilst in the ideal NLS model they stay decoupled. Commonly, such deviations are due to fibre loss and amplification noise; the latter is usually modelled as additive white Gaussian noise (AWGN) \cite{alb18, tpl17}. We studied whether the presence of the bound states influences the system's response to noise in the time domain. To make the comparison clear, we introduce two separate quantities, SNR$_b$ and SNR$_q$, as measures of the noise affecting the functions $b(k)$ and $q(t)$, correspondingly. Starting from the function $b(k)$ and optionally embedding the additional soliton, we then compute the inverse NFT to find the optical field waveform, add AWGN in the time domain, and use the direct NFT to evaluate the $b(k)$ of the noisy signal. Then we compare the initial and back-computed $b(k)$ to evaluate the value of SNR$_b$. Note that to make the comparison fair, we count the signal power for the SNR$_q$ definition as it would be \emph{in the absence of the solitonic eigenvalue}. We identify the qualitative difference depending on the signal amplitude, see Fig.~\ref{fig:snrs}.
}

\red{For the small-amplitude case ($A=0.2$, left panel in Fig.~\ref{fig:snrs}), we observe a slight deterioration of the SNR for the back-computed $b(k)$, if the signal contains the solitonic mode. However, we see that for the larger amplitude ($A=0.7$, right panel in Fig.~\ref{fig:snrs}), the value of the SNR for the soliton-free profile is smaller than that in the presence of the additional soliton mode. As far as the higher amplitudes are more relevant for NFT-based transmission where NFT can have an advantage over conventional modulation, so we can conclude from our test that the introduction of the solitonic modes may even potentially improve the overall transmission quality.}

\red{Another issue in question can be the growth of the peak-to-average power ratio (PAPR) in the presence of eigenvalues (e.g. for profiles in Fig.~\ref{fig:sincs}a, PAPR of the purely $b$-modulated waveform is 6.7 dB, whilst additions of the solitonic modes increase it up to 11.8 dB and 13.5 dB, respectively). This may cause some undesired transmitter-induced nonlinear effects. However, we would like to note that the procedure described in this paper allows us to embed eigenmodes from an infinite set. So one can pick several solitonic modes with specially chosen parameters to generate signals with the lower PAPR.
}

\section{Conclusion}
In this work, we filled the gap in the rigorous mathematical formulation of the $b$-modulation method, constituting the most efficient up to date technique within the NFT-based communications. We presented the explicit proofs providing the one-to-one correspondence between the nonlinear spectrum, which satisfies the requirements of the $b$-modulation, and the optical signal with finite pre-defined time support. In addition, we presented the full procedure and the mathematical proofs related to the important open question: how to implant the discrete solitary modes into the $b$-modulation concept without violating the condition of the exact localisation of the time-domain profile. Our results were eventually illustrated and satisfactory validated through direct numerical analysis. \red{The additional solitonic modes can provide more flexibility for the design of $b$-modulated long-haul optical transmission systems, even though the additional modes cannot directly render us the additional parameters for the modulation (aside from adding those modes in an on-off keying way) if we aim at keeping the exact localisation. There also emerges an interesting question of whether the  solitonic modes embedded in the $b$-modulation can be used as additional data carriers if we relax the exact localisation constraint (as was done in \cite{yal18,yal19,yla19}) and/or whether they can be used for the improvement of system's performance}.

\section*{Acknowledgement}
We would like to thank Dr Nikolay Chichkov for help with the manuscript preparation and the interest in this work.

\end{document}